\documentclass{aa}
\usepackage[english]{babel}%
\usepackage{graphicx}%
\usepackage{amssymb}
\usepackage{hhline}
\usepackage{tablefootnote}
\newlength{\deftabcolsep}

\begin{document}

\title{On structure and kinematics of the Virgo cluster of galaxies\thanks{%
Table~\ref{kkk19:tab01} is available at the CDS via anonymous ftp to
cdsarc.u-strasbg.fr (130.79.128.5) or via
http://cdsarc.u-strasbg.fr/viz-bin/cat/J/A+A/}}

\titlerunning{Virgo cluster of galaxies} 

\author{Olga G. Kashibadze\inst{1}\fnmsep\thanks{f.k.a. Nasonova.},
        Igor D. Karachentsev\inst{1},
   \and Valentina E. Karachentseva\inst{2}}
\authorrunning{O.\,Kashibadze, I.\,Karachentsev, \& V.\,Karachentseva}     
        
\institute{Special Astrophysical Observatory of the Russian Academy of Sciences\\
          Nizhnij Arkhyz, Karachay-Cherkessia, 369167, Russia\\
\and
          Main Astronomical Observatory of the National Academy of Sciences\\
          27 Akademika Zabolotnoho St., Kyiv, 03143, Ukraine}

\date{\today}

\abstract
{}
{This work considers the Virgo cluster of galaxies, focusing on its structure,
kinematics, and morphological landscape. Our principal aim is to estimate the
virial mass of the cluster. For this purpose, we present a sample of 1537
galaxies with radial velocities $V_{LG} < 2600$~km~s$^{-1}$ situated within a
region of $\Delta{}SGL = 30^\circ$ and $\Delta{}SGB = 20^\circ$ around M87. 
About half of the galaxies have distance estimates.}
{We selected 398 galaxies with distances in a $(17\pm5)$~Mpc range. Based on
their 1D and 2D number-density profiles and their radial velocity dispersions, we
made an estimate for the virial mass of the Virgo cluster.}
{We identify the infall of galaxies towards the Virgo cluster core along the
Virgo Southern Extension filament. From a 1D profile of the cluster, we obtain
the virial mass estimate of $(6.3\pm0.9) \times 10^{14} M_\odot, $ which is in tight
agreement with its mass estimate via the external infall pattern of galaxies.}
{We conclude that the Virgo cluster outskirts between the virial radius and the
zero-velocity radius do not contain significant amounts of dark matter beyond
the virial radius.}

\keywords{galaxies: clusters: individual: Virgo}

\maketitle

\section{Introduction}
It is well known that a thousand of the brightest galaxies with apparent
magnitudes $m_B < 13^m$ (Sandage \& Tammann 1981) form a planar structure across
the sky, called the Local Supercluster (de Vaucouleurs 1961). The most dense
region of the Local Supercluster (LSC) is the Virgo cluster of galaxies, with
the centre usually identified by the radiogalaxy Virgo\,A$=$M87$=$NGC\,4486.

In our previous paper (Kashibadze et al. 2018$\,=\,$KKK18), we considered the
distribution and the radial velocity field of galaxies located within the
equatorial belt of 100 by 20 degrees, centred on M87 with its supergalactic
coordinates $SGL = 102.88^\circ$ and $SGB = -2.35^\circ$. This region, including
the whole Virgo cluster and its spurs, contains 2158 galaxies with radial
velocities $V_{LG}<2000$~km~s$^{-1}$ relative to the Local group centroid. Half
of them have distance estimates $D$ and peculiar velocities $V_{pec} =
V_{LG} - H_0 \times D$, where $H_0$ is the local Hubble parameter, which is
set to 73~km~s$^{-1}$~Mpc$^{-1}$. It has been found out that some galaxy groups
in the belt have peculiar velocities up to $~\sim(500-800)$~km~s$^{-1, }$
comparable with virial velocities in rich clusters of galaxies. Using the
spherically symmetrical infall model, we estimated the radius of the
zero-velocity surface of the Virgo cluster as $23^\circ,$ or $(7.8\pm0.3)$~Mpc,
corresponding to the total mass of the cluster $M_T = (7.4\pm0.9) \times 10^{14}
M_\odot$. Shaya et al. (2017) obtained nearly the same value of the radius of
the zero-velocity surface, $(7.3\pm0.3)$~Mpc, considering trajectories of
galaxies within the Numerical Action Method.

In this work, which is a  continuation of the previous KKK18 study,
we restrict our analysis with the region limited by $\Delta{}SGL =
30^\circ$ and $\Delta{}SGB = 20^\circ$ centred at M87. The sample used in KKK18
was supplemented by galaxies with radial velocities $V_{LG}
=[2000-2600]$~km~s$^{-1}$ and new distances from Cantiello et al. (2018) and
Toloba et al. (2018). 

It should be noted that the Virgo Cluster Catalog ($=$VCC, Binggeli et al.
1985) and the Extended Virgo Cluster Catalog ($=$EVCC, Kim et al. 2014) have
made a foundational contribution to studies of the structure of the Virgo
cluster. Both catalogues have expanded our insights into the dwarf population of
the nearest cluster of galaxies. Still, there are no measured radial velocities
for most dwarf galaxies in the Virgo region, and their membership in the cluster
is considered from some indirect indicators. Tonry et al. (2000) and Mei et al.
(2007) performed bulk distance measurements for members of the Virgo
cluster. The surface-brightness-fluctuation method they had used provided
distances for more than a hundred of the most bright E and S0 galaxies with an
accuracy of $\sim10-15$\,\%. The data are very important for understanding the
structure and the dynamics of the Virgo cluster, since radial velocities of
galaxies in the vicinity of a massive attractor are ambiguous and uncertain
indicators of their distances.

 The ALFALFA $HI$-survey of galaxies in Arecibo (Haynes et al. 2011, 2018)
 brought radial velocities and $HI$-line widths for many late-type galaxies in
 the Virgo cluster area. This allowed us to estimate new distances for 380
 irregular and spiral galaxies in the sky region in question, doubling the total
 number of galaxies with known distances. This progress in distance estimates
 provided us with an opportunity to review the structure and dynamics of the
 Virgo cluster again.  In total, we have a sample of 1537 galaxies inside the
 area of $30^\circ$ by $20^\circ$ with measured radial velocities. Among these
 galaxies, 738 (48\,\%) have distance estimates.

The paper is organised as follows: In Section 2, we present the sample selected
for our analysis. The total luminosity of the Virgo cluster core in $B$ and $K$
bands is briefly summarised in Section 3. Distributions of early-type (E, S0,
dSph), spiral (Sa-Sm) and irregular (I, Im, BCD) galaxies on the sky and in the
depth are presented in Section 4. Section 5 describes 2D and 1D profiles of the
Virgo cluster delineated by galaxies within a distance range of ($17\pm5$) Mpc,
and it presents various estimates of the virial mass of the cluster. The basic
parameters of the Virgo cluster are fixed in Section 6. A~brief summary of our
results are given in Section 7. A Virgo cluster distance of 16.5 Mpc (Mei et al.
2007) is used throughout the paper, where $6^\circ = 1.7$ Mpc is the cluster
virial radius.

\section{The Virgo cluster core and suburbs}
\begin{sidewaystable*}
\caption{Galaxies around M87 with supergalactic coordinates SGL=[87.9,
117.9]$^{\circ}$, SGB=[-12.4,+7.6]$^{\circ},$ and radial velocities $V_{LG} <
2600$ km~s$^{-1}$.}\label{kkk19:tab01}
{\normalsize\begin{tabular}{lcrrrrlrrrrrrl}\hline\hline
Name &
\multicolumn{1}{c}{RA (2000) Dec} &
\multicolumn{1}{c}{SGL} &
\multicolumn{1}{c}{SGB} &
\multicolumn{1}{c}{$R_p$} &
\multicolumn{1}{c}{$V_{LG}$} &
\multicolumn{1}{c}{Type} &
\multicolumn{1}{c}{$B_T$} &
\multicolumn{1}{c}{$K$} &
\multicolumn{1}{c}{$W_{50}$} &
\multicolumn{1}{c}{$m_{21}$} &
\multicolumn{1}{c}{$m-M$} &
\multicolumn{1}{c}{$D$} &
meth\\
&&
\multicolumn{1}{c}{($^{\circ}$)}&
\multicolumn{1}{c}{($^{\circ}$)}&
\multicolumn{1}{c}{($^{\circ}$)}&
\multicolumn{1}{c}{(km~s$^{-1}$)}&&
\multicolumn{1}{c}{(mag)}&
\multicolumn{1}{c}{(mag)}&
\multicolumn{1}{c}{(km~s$^{-1}$)}&
\multicolumn{1}{c}{(mag)}&
\multicolumn{1}{c}{(mag)}&
\multicolumn{1}{c}{(Mpc)}&\\
\hline
PGC1664006    & J113206.6+220731 & 88.93 &$-$12.33 &17.05 &1091 &Im  &17.36 &15.01 & 47 &18.08 &31.28 &18.00 &TF   \\  
PGC1620526    & J113554.4+201320 & 91.08 &$-$12.24 &15.31 &1038 &I   &17.04 &14.69 &108 &17.69 &  ... &  ... &     \\  
PGC1597887    & J113908.9+193460 & 91.97 &$-$11.77 &14.34 &1636 &I   &18.28 &15.93 &... &  ... &  ... &  ... &     \\  
UGC06670      & J114229.4+181959 & 93.45 &$-$11.49 &13.08 & 823 &I   &13.56 &11.21 &194 &14.60 &31.29 &18.10 &TF   \\  
PGC1519757    & J114440.7+165359 & 95.01 &$-$11.51 &12.04 & 813 &BCD &16.98 &14.63 & 91 &17.97 &  ... &  ... &     \\  
UGC06747      & J114624.0+134938 & 98.11 &$-$12.18 &10.91 &2574 &Sdm &16.27 &13.67 & 96 &15.30 &31.53 &20.30 &TFb  \\ 
EVCC2007      & J114626.5+193012 & 92.69 &$-$10.19 &12.80 &2086 &Im  &19.16 &16.81 &102 &17.66 &33.78 &57.10 &TFb  \\ 
AGC219628     & J114714.0+184028 & 93.54 &$-$10.31 &12.22 &1746 &Sm  &18.50 &16.15 & 77 &17.49 &32.73 &35.10 &TFb  \\ 
SDSSJ114717   & J114717.5+162005 & 95.77 &$-$11.12 &11.26 & 700 &E   &17.97 &13.87 &... &  ... &  ... &  ... &     \\  
PGC2806928    & J114816.4+183833 & 93.65 &$-$10.09 &12.00 & 876 &Im  &16.64 &14.29 & 65 &16.41 &31.31 &18.24 &TF   \\  
SDSSJ114843   & J114843.1+171053 & 95.08 &$-$10.50 &11.24 & 963 &I   &17.38 &15.03 & 42 &17.72 &30.92 &15.31 &TF   \\  
AGC219539     & J114845.0+164423 & 95.51 &$-$10.65 &11.07 &1510 &I   &18.60 &16.25 & 72 &17.18 &32.64 &33.85 &TFb  \\ 
DDO097        & J114857.4+235016 & 88.79 &$-$8.09  &15.16 & 452 &I   &15.12 &12.77 & 86 &15.50 &30.69 &13.74 &bs   \\  
PGC1528400    & J114905.6+171520 & 95.04 &$-$10.39 &11.19 & 519 &BCD &18.06 &15.71 &... &  ... &  ... &  ... &     \\ 
AGC219541     & J114921.7+171724 & 95.03 &$-$10.32 &11.15 &1266 &Im  &18.20 &15.85 & 66 &17.50 &32.55 &32.33 &TFb  \\ 
SDSSJ114931   & J114931.0+151540 & 96.99 &$-$10.98 &10.42 & 747 &E   &17.85 &13.75 &... &  ... &  ... &  ... &     \\  
SDSSJ114957   & J114957.1+161744 & 96.03 &$-$10.53 &10.64 &1110 &I   &18.11 &15.76 & 33 &18.34 &30.89 &15.10 &TF   \\  
PGC036976     & J115002.7+150124 & 97.26 &$-$10.94 &10.24 & 638 &BCD &15.43 &13.08 & 47 &16.80 &30.67 &13.60 &TF   \\  
PGC037048     & J115055.9+143542 & 97.74 &$-$10.88 & 9.94 & 893 &Im  &15.52 &13.17 & 52 &15.70 &30.32 &11.60 &TF   \\  
KIG0506       & J115201.9+135244 & 98.52 &$-$10.87 & 9.56 & 847 &I   &16.12 &13.77 & 79 &16.61 &31.31 &18.30 &TF   \\  
SDSSJ115220   & J115220.2+152736 & 97.03 &$-$10.27 & 9.82 & 676 &Im  &17.53 &15.18 & 68 &17.65 &  ... &  ... &     \\  
SDSSJ115300   & J115300.3+160230 & 96.53 &$-$9.93  & 9.86 & 823 &I   &18.53 &16.18 & 23 &19.93 &  ... &  ... &     \\  
KK111         & J115401.6+164323 & 95.96 &$-$9.47  & 9.90 & 879 &I   &17.06 &14.71 & 41 &16.77 &30.32 &11.59 &TFb  \\  
AGC215145     & J115412.5+122606 &100.09 &$-$10.84 & 8.93 & 880 &I   &19.00 &16.65 & 32 &16.80 &29.57 & 8.20 &TFb  \\ 
UGC06881      & J115444.9+200323 & 92.85 &$-$8.17  &11.56 & 522 &I   &15.83 &13.48 & 90 &16.20 &31.07 &16.40 &TF   \\ 
NGC3976       & J115557.4+064459 &105.74 &$-$12.22 &10.27 &2348 &Sb  &12.45 & 8.60 &436 &13.27 &32.53 &32.70 &tf   \\ 
PGC037490     & J115602.5+064041 &105.81 &$-$12.23 &10.30 &2354 &S0  &15.83 &11.73 &... &  ... &  ... &  ... &     \\  
AGC215716     & J115707.2+064032 &105.90 &$-$11.97 &10.08 &2358 &Sm  &17.70 &15.35 &100 &17.39 &33.75 &56.20 &TF   \\  
PGC1488625    & J115840.4+153534 & 97.42 &$-$8.79  & 8.43 & 458 &I   &18.28 &15.93 &... &  ... &  ... &  ... &     \\  
AGC213178     & J115900.7+044011 &108.00 &$-$12.12 &11.01 &1451 &BCD &16.91 &14.56 & 58 &18.20 &31.44 &19.40 &TF   \\  
PGC037779     & J115933.8+135315 & 99.12 &$-$9.14  & 7.75 &1334 &Sm  &15.09 &12.74 & 57 &15.70 &30.27 &11.30 &TF   \\  
PGC037802     & J115949.4+212656 & 91.94 &$-$6.58  &11.70 &1893 &Sm  &15.81 &13.46 &124 &16.02 &32.77 &35.81 &TF   \\  
NGC4032       & J120032.9+200426 & 93.30 &$-$6.88  &10.57 &1186 &Sb  &13.03 & 9.18 &107 &14.28 &  ... &  ... &     \\  
PGC1380373    & J120056.1+102956 &102.49 &$-$9.90  & 7.56 &2529 &Im  &16.44 &14.09 & 83 &16.84 &32.12 &26.50 &TF   \\ 
IC0755        & J120110.5+140615 & 99.04 &$-$8.70  & 7.41 &1419 &Sbc &14.09 &10.49 &206 &14.73 &32.34 &29.30 &tf   \\
\hline
\end{tabular}}
\footnotetext{The table contains 1537 galaxies with radial velocities $V_{LG} <
2600$~km~s$^{-1}$. Among them, 738 galaxies have distance estimates: 358 from
public databases and 380 distances estimated for the first time in this paper
(indicated in the last column as TF or TFb).}
\end{sidewaystable*}

The basic observational characteristics of 1537 galaxies in the considered area
are presented in Table~\ref{kkk19:tab01}, with the full machine readable version
available at SIMBAD Astronomical Database\footnote{http://simbad.u-strasbg.fr/simbad/}.
The table columns contain the (1) galaxy name; (2) equatorial coordinates J2000.0;
(3, 4) supergalactic coordinates; (5) projected distance of a galaxy from M87;
(6) radial velocity (km~s$^{-1}$) relative to the Local Group centroid; (7)
morphological type determined by us based on images from Sloan Digital Sky
Survey (SDSS, Abazajian et al. 2009) and Panoramic Survey Telescope and Rapid
Response System (PanSTARRS, Chambers et al. 2016); (8) apparent $B$-magnitude
from HyperLEDA\footnote{http://leda.univ-lyon1.fr/} (Makarov et al. 2014) or
NASA Extragalactic Database\footnote{http://ned.ipac.caltech.edu/} (NED); (9)
apparent $K$-magnitude calculated from $B$-magnitude adjusted for the morphological
type of the galaxy according to Jarrett et al. (2003); (10) 21~cm $HI$ line width
(km~s$^{-1}$) at half maximum from HyperLEDA data; (11) apparent $HI$-magnitude
from HyperLEDA; (12, 13) distance modulus and galaxy distance (Mpc), (14) method
applied to determine distance: `cep', `SN'--from supernovae and cepheids
luminosity, `sbf'--from surface brightness fluctuation, `rgb'--from the tip of
the red giant branch, `bs'--from the brightest stars luminosity, `FP'--from the
fundamental plane, `gc'--from luminosity function of globular clusters, `tf',
`TF', `TFb'--from Tully \& Fisher relation (1977), with lowercase indicating
distance estimates from NED, and uppercase indicating our estimates based on
classical (Tully \& Pierce 2000) or baryonic (Karachentsev et al. 2017)
Tully-Fisher relation.  

\begin{figure} 
\centering
\includegraphics[width=\hsize]{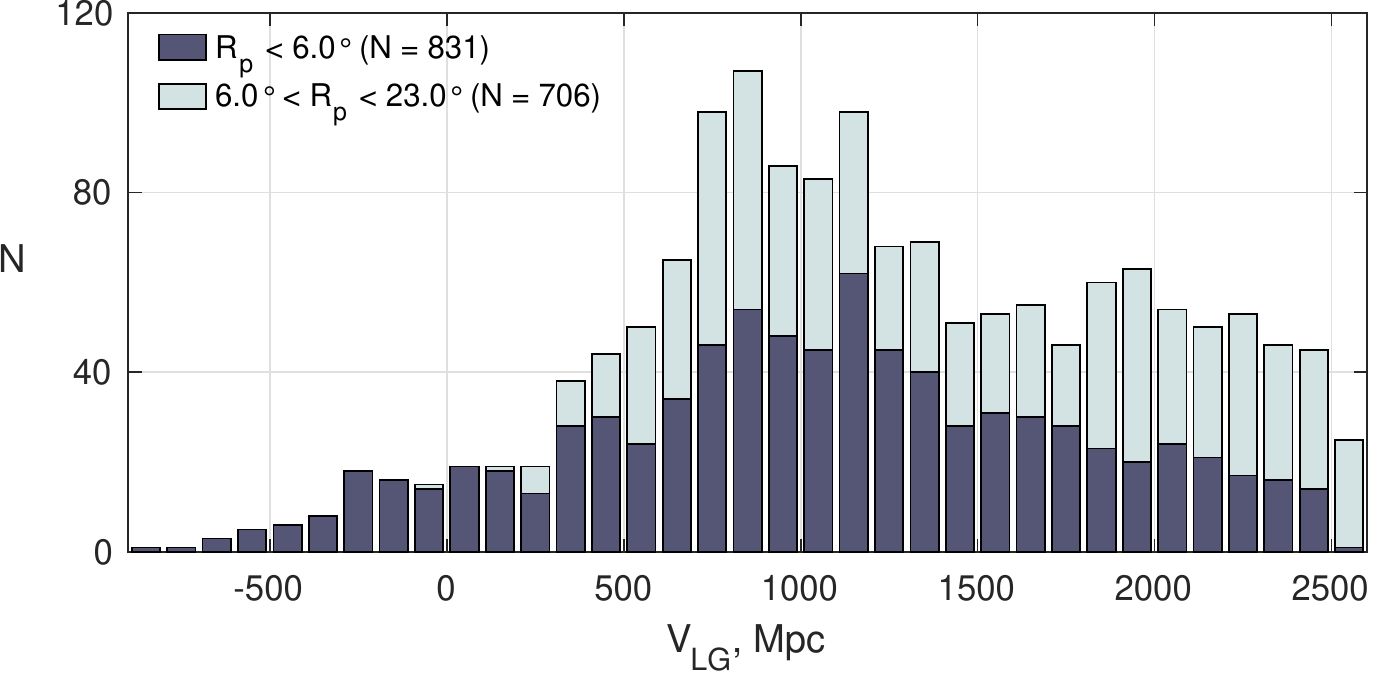}
\caption{Distribution of galaxies  in Virgo cluster area by their 
radial velocities relative to
the Local Group centroid. The virial core members are shown with a darker
colour.}
\label{kkk19:fig01}
\end{figure}

The distribution of galaxies in the considered area by their radial velocities
relative to the Local Group centroid is presented in Fig.~\ref{kkk19:fig01}.
Galaxies populating the central virial zone of the cluster limited by $6.0^\circ$
are marked with a darker colour. Clearly, the members of the virial core
of the cluster show quite a symmetrical distribution around the mean velocity of
$\sim1000$~km~s$^{-1}$. The region of negative radial velocities is populated
entirely by the virial core galaxies, providing evidence of nearly radial
motions (Karachentsev \& Kashibadze 2010). The relative number of galaxies in
the outskirts of the cluster grows with higher values of $V_{LG}$ and becomes
predominant with $V_{LG} > 1800$~km~s$^{-1}$. Assuming the symmetry of
the velocity distribution relative to the average, we estimate that
$1-2$\,\% of the virial core
members may lie beyond $V_{LG} > 2600$~km~s$^{-1}$, being lost among field
galaxies.

\begin{figure} 
\centering
\includegraphics[width=\hsize]{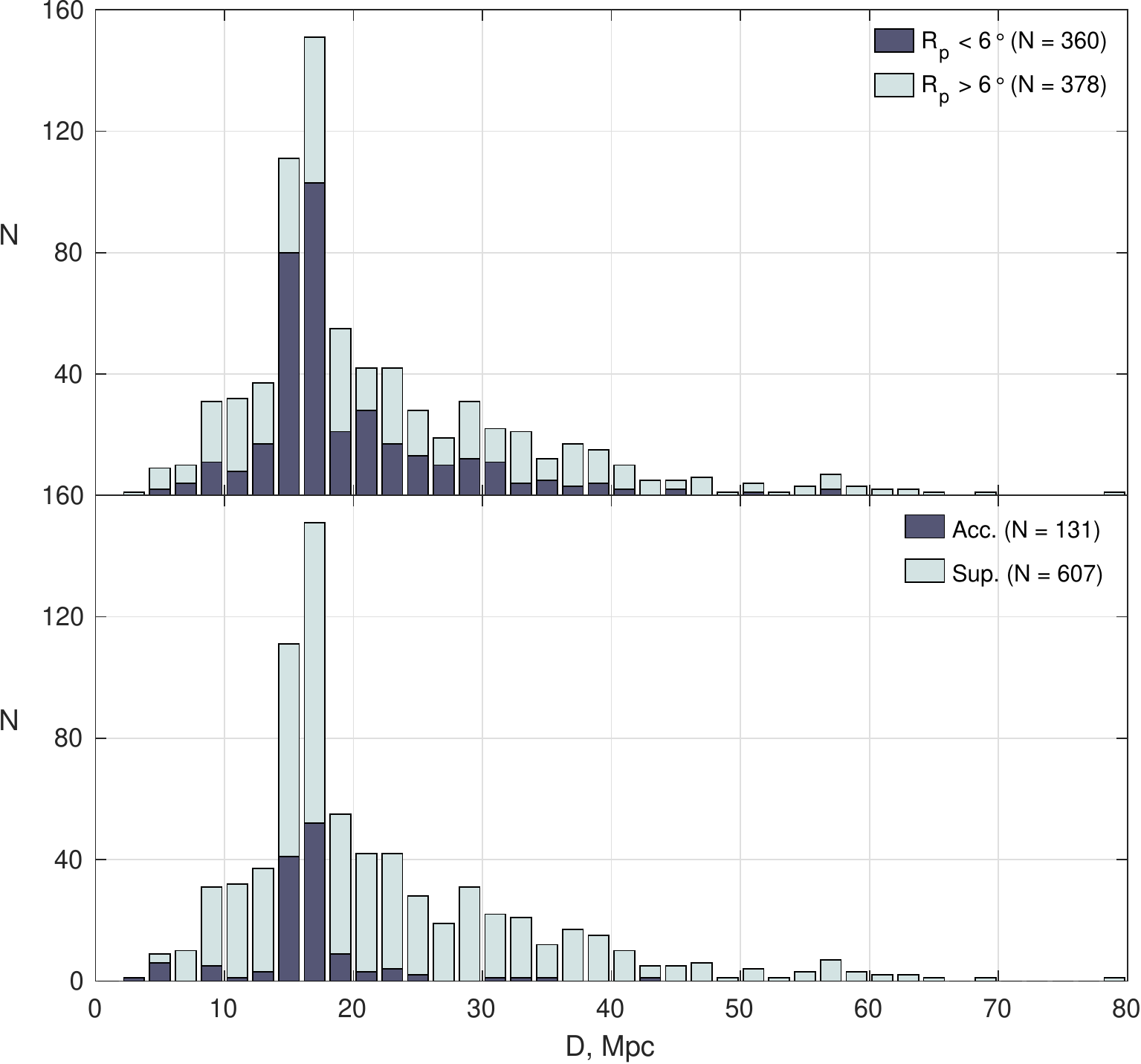}
\caption{Distribution of galaxies in Virgo cluster area by their 
distances. Upper panel: the
galaxies of the virial core are marked with a darker colour. Lower panel: the
galaxies with `accurate' distances (sbf, cep, SN, rgb) are shown with a darker
colour, while those with less reliable (`supplementary') distances (tf, FP, bs,
gc) are shown with a lighter colour.}
\label{kkk19:fig02}
\end{figure}

The upper panel of Fig.~\ref{kkk19:fig02} reproduces the distribution of
galaxies in our area by distances. The galaxies of the virial core are marked
with a darker colour. About a half of the virial zone galaxies have distance
estimates lying within $(14-18)$~Mpc, suggesting their real membership in the
Virgo cluster with its mean distance of $16.5\pm0.2$~Mpc according to Mei et al.
(2007). Peripheral galaxies also show the peak at $(14-20)$~Mpc, though far less 
distinctly. The significant amount of galaxies resides at distances $D > 30$~Mpc. 
Their presence in the $V_{LG} < 2600$~km~s$^{-1}$ sample is partially caused by
infall of background galaxies towards the centre of the attractor, as well as by
the existence of a large expanding cosmic void between the Virgo and the Coma
clusters, which imparts an additional velocity component towards an observer.
Some galaxies in the larger $D$ tail area may be caught due to errors in
distance estimates.

We combined our data into two groups: the first containing galaxies with
accurately measured distances (sbf, cep, SN, rgb), and the second including
galaxies with less reliable distances (tf, FP, bs, gc). The typical distance
error is about 10\,\% in the first sub-sample, and about 25\,\% in the second one.
The distribution of galaxies of these two categories by distances is shown in
the lower panel of the Fig.~\ref{kkk19:fig02}. About $3/4$ of galaxies with
accurate distances lie in a virial interval of $(14-18)$~Mpc, while this rate is
only about $1/4$ among galaxies with distances estimated from secondary
indicators. It should be noted that this difference is caused not only by the
order of distance errors, but also by the morphological segregation: the
surface-brightness-fluctuation method is applied to E, S0, and dSph galaxies populating the
centre of the cluster, while Tully-Fisher relation is suitable for gas-rich S,
I, and BCD galaxies inhabiting some less dense regions.

\begin{figure*} 
\centering
\resizebox{0.7\hsize}{!}
{\includegraphics{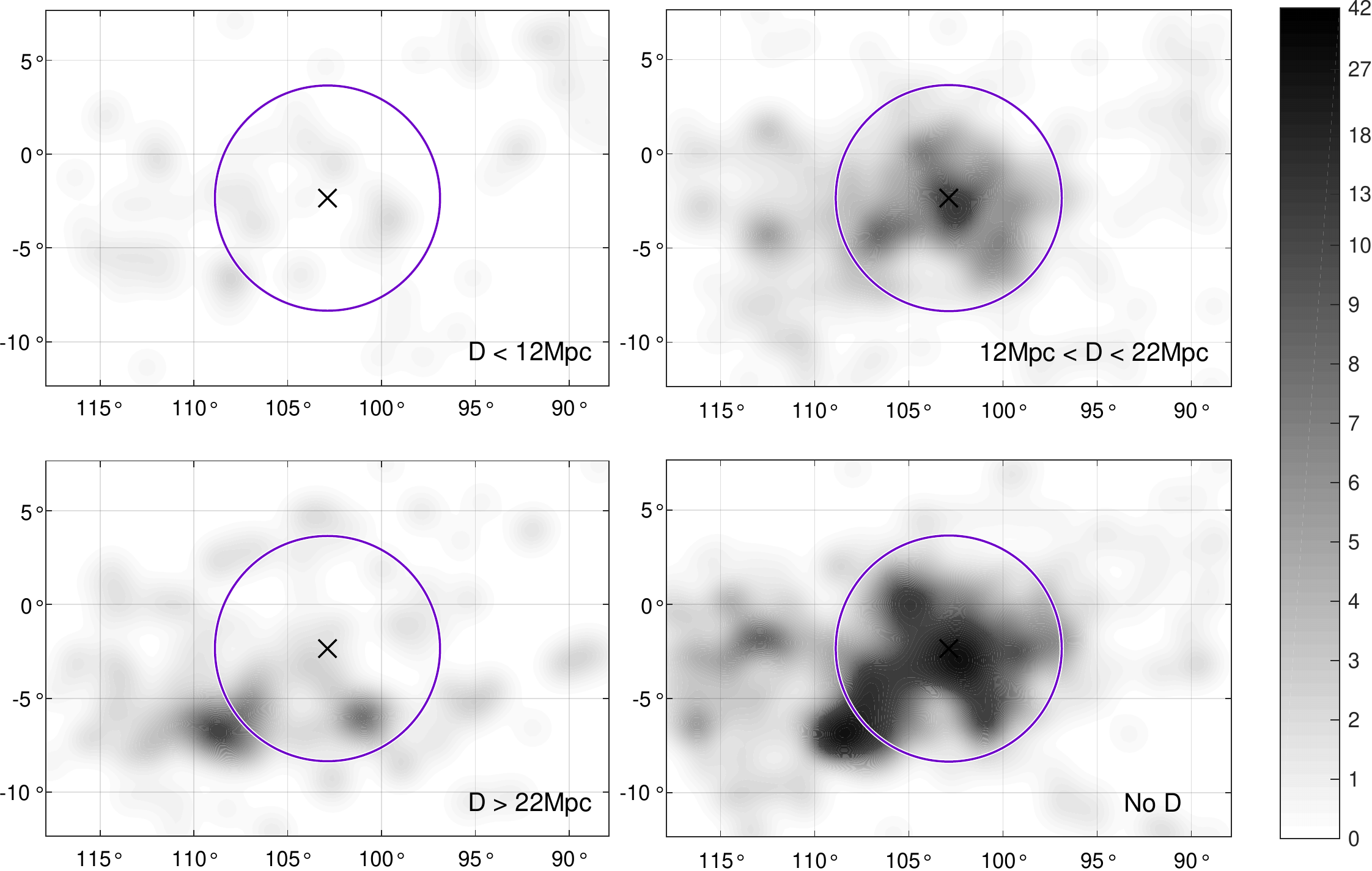}}
\caption{Sky distribution of galaxies  in Virgo cluster area within 
different distance intervals.
The smoothing window is $0.75^\circ$. The circle corresponds to the virial radius
of $6^\circ$.}
\label{kkk19:fig03}
\end{figure*}


The distribution of galaxies within different distance intervals over the sky is
presented in Fig.~\ref{kkk19:fig03}. We enlarged the interval
containing Virgo members up to $(12-22)$~Mpc to account not only for virial
depth of the cluster ($\pm2$~Mpc), but also for distance errors inherent in
the Tully-Fisher method ($\pm4$~Mpc). The surface density distribution is smoothed
with a window of $0.75^\circ$. The circle corresponds to the virial radius of
$6^\circ$. As can be seen from these data, the foreground galaxies in the
upper-left panel do not demonstrate any clumping towards the cluster centre.
Most galaxies with distances falling into $(17\pm5)$~Mpc are located in the sky
within the virial radius. The background galaxies with $ D > 22$~Mpc group along
the filament, which includes several clumps (Kim et al. 2016), the most rich of
them are $W$ and $M$ clouds according to de Vaucouleurs. The main galaxies in
these clouds (Makarov \& Karachentsev 2011), their coordinates, the mean radial
velocities and distances are listed in Table~\ref{kkk19:tab02}.

\setlength{\deftabcolsep}{\tabcolsep}
\setlength{\tabcolsep}{4pt}
\begin{table}
\centering
\caption{Galaxy clouds behind the Virgo cluster.}\label{kkk19:tab02}
\begin{tabular}{ccccrcc}\hline\hline
Cloud&Galaxy&SGL&SGB&$N_V$&$\langle V_{LG}\rangle$&$\langle D \rangle$\\
&&($^{\circ}$)&($^{\circ}$)&&(km~s$^{-1}$)&(Mpc)\\
\hline
W     &NGC4261 & 108.4 & -6.9 &  87   & 2060 &  29.4\\
M     &NGC4189 & 100.7 & -6.0 &   6   & 1987 &  30.0\\
\hline
\end{tabular}
\end{table}
\setlength{\tabcolsep}{\deftabcolsep}

At present, half of the galaxies in the considered area with $V_{LG} <
2600$~km~s$^{-1}$ have not distance estimates. These are predominantly
intermediate- and low-luminosity galaxies. As we can see from the bottom-right
panel of Fig.~\ref{kkk19:fig03}, about 60\,\% of these galaxies are inside the
virial radius of the Virgo cluster, and the majority of the rest belong to the
background $W$ cloud.

\section{Luminosity of the Virgo cluster}
The determination of the total luminosity of the Virgo cluster in $B$ or $K$
band is a non-trivial task, because it requires an accurate knowledge of
cluster membership. The SDSS spectral survey Abazajian et al. 2009 
and the ALFALFA survey in
21~cm $HI$ line (Haynes et al. 2018) have significantly enriched the sample of
galaxies with measured radial velocities in the Virgo region. The reasonable
completeness of surveys in radial velocities has already been achieved for
$B\simeq16^m-17^m$ corresponding to luminosity of a dwarf galaxy ($M_B \simeq
-15^m - 14^m$) at the Virgo distance. However, the radial velocity can't be a
reliable indicator of distance in the vicinity of a massive attractor.
Morphological type and angular distance from the cluster centre are usually
regarded as additional markers, confirming that galaxies are members of a cluster.

\begin{figure} 
\centering
\includegraphics[width=\hsize]{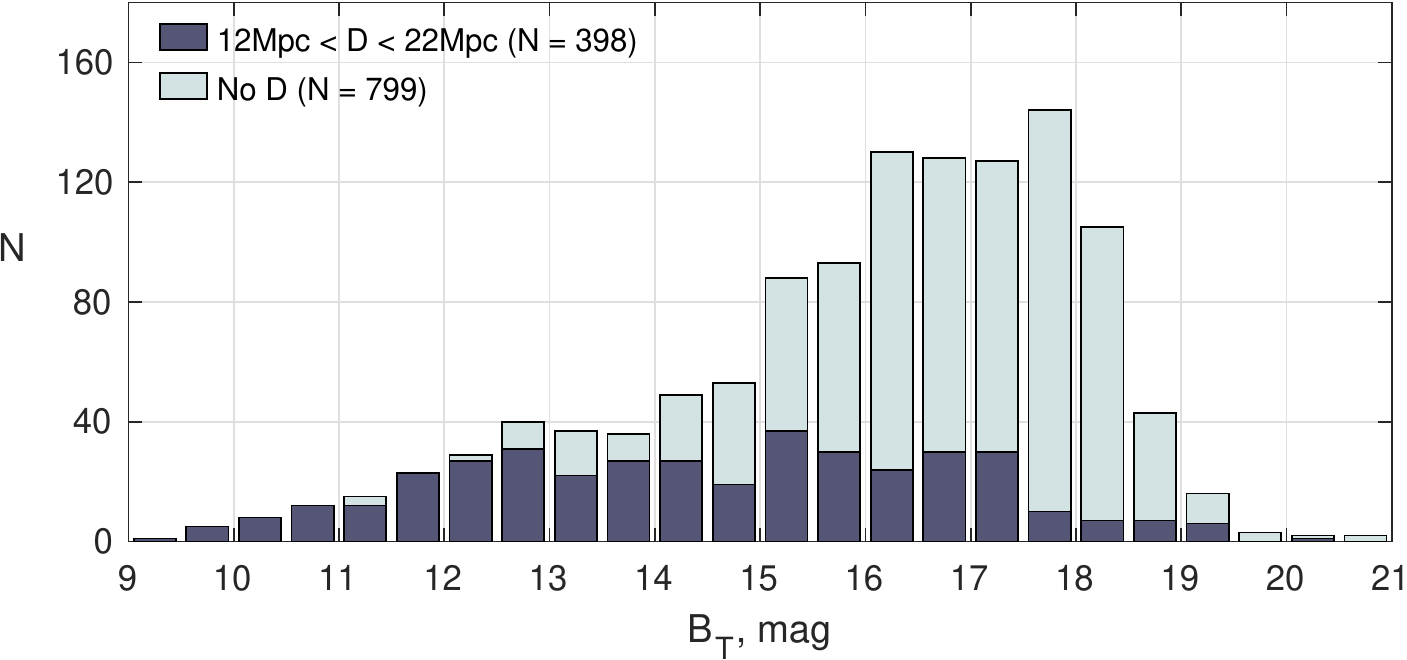}
\caption{Distribution of galaxies in Virgo cluster area with 
measured radial velocities by the apparent magnitude $B_T$. 
The galaxies with distances $D = (12-22)$~Mpc are shown with
a darker colour, while those with no distances are shown with a lighter colour.}
\label{kkk19:fig04}
\end{figure}

\begin{figure*} 
\centering
\resizebox{0.7\hsize}{!}
{\includegraphics{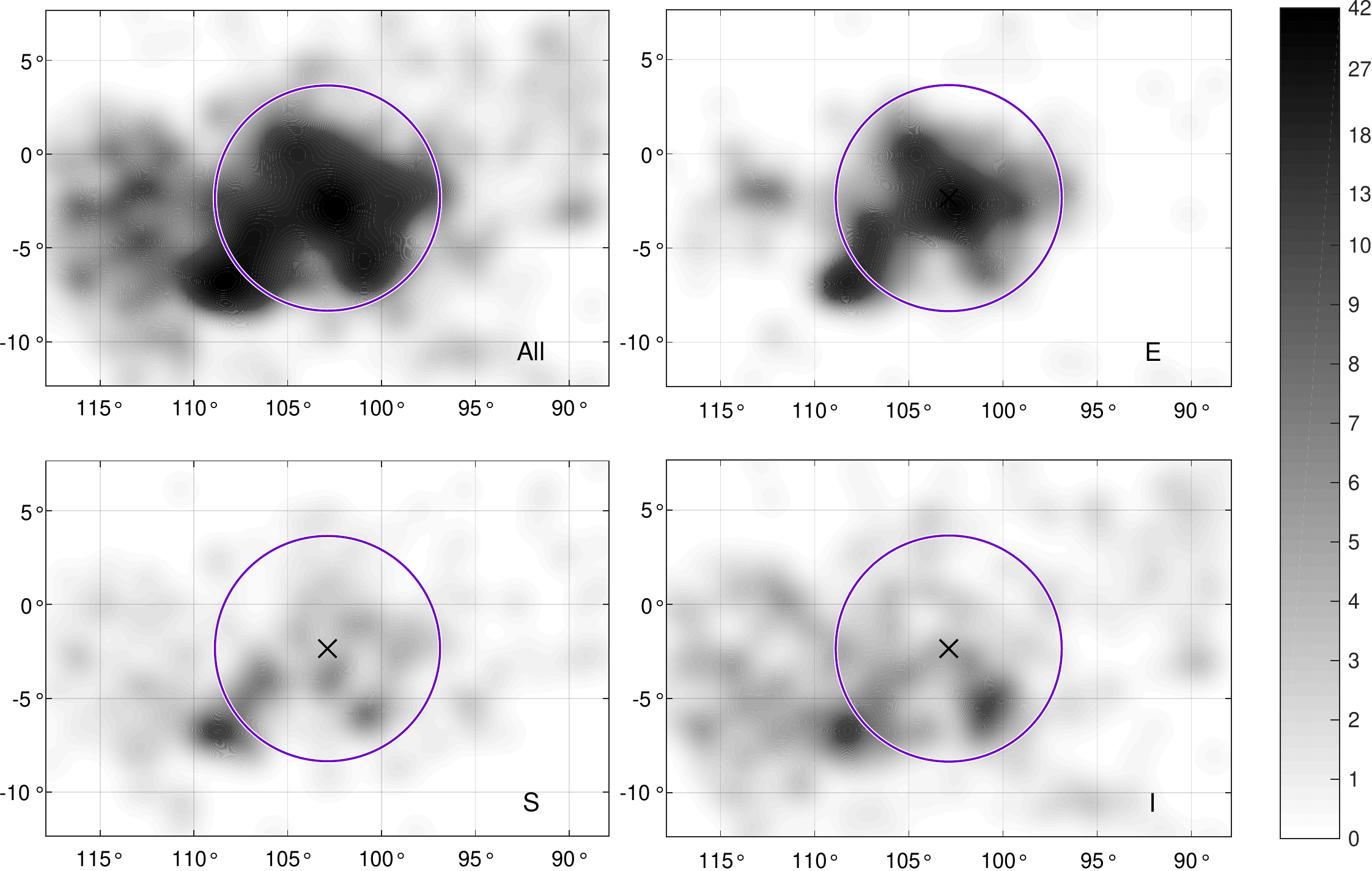}}
\caption{Sky distribution of galaxies of different morphological types
 with $V_{LG} < 2600$ km/s in the Virgo cluster area.
The smoothing window is $0.75^\circ$. The circle corresponds to the virial radius
of $6^\circ$.}
\label{kkk19:fig05}
\end{figure*}


Figure~\ref{kkk19:fig04} represents the distribution of galaxies with distances $D
= (12-22)$~Mpc by $B_T$. The lighter colour indicates galaxies from
Table~\ref{kkk19:tab01} with distances that have not yet been measured. As one
can see, about $90$\,\% of the galaxies with $ B_T < 13 $ mag have distance
estimates. These bright galaxies with $M_B < -18^m$ make a prevailing
contribution to the integrated luminosity of the cluster. We estimated the
total luminosity of the galaxies with $D = (12-22)$~Mpc located inside the
$6^\circ$ radius to be $L_B = (1.8\pm0.2) \times 10^{12}L_\odot$. Assuming the
Schechter luminosity function with the slope parameter $\alpha = -1.3$ (Sandage
et al. 1985), we estimated that about 25\,\% of this quantity is contributed by
faint galaxies with as yet unmeasured distances. Therefore, we fix $1.8 \times
10^{12}L_\odot$ as an estimate of blue luminosity of the Virgo cluster core. The
similar calculations applied to the distribution of galaxies by their
$K$-magnitudes yield the total luminosity of the cluster core as $L_K =
(8.6\pm1.1) \times 10^{12}L_\odot$. These quantities can be compared with the
data from literature. Thus, Sandage et al. (1985) derived the total blue
luminosity of the Virgo galaxies within $6^\circ$ core to be $1.4 \times
10^{12}L_\odot$ at the distance of 16.5 Mpc. According to Kourkchi and Tully
(2017), the galaxies in the whole Virgo association have an integrated $K$-band
luminosity of $1 \times 10^{13}L_\odot,$ with $5 \times 10^{12}L_\odot$ in the
collapsed core.

\begin{figure} 
\centering
\includegraphics[width=\hsize]{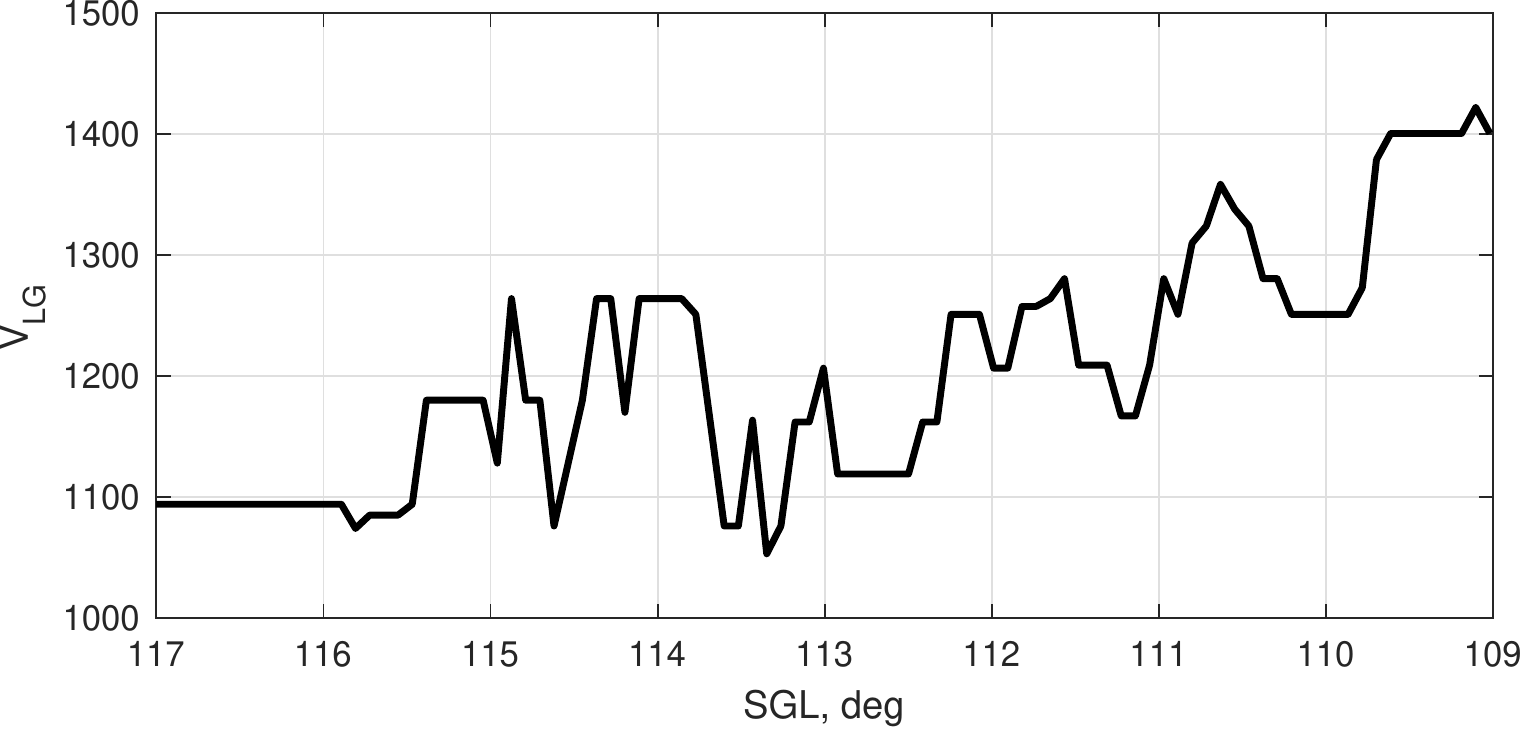}
\caption{Radial velocities of galaxies in Virgo Southern Extension as a
function of supergalactic longitude. The broken line represents the running
median with a window of $2^\circ$.}
\label{kkk19:fig06}
\end{figure}

\section{Morphological structure}
The relation between morphology and kinematics of galaxies in the Virgo cluster
has been studied by de Vaucouleurs \& de Vaucouleurs (1973),
Tully \& Shaya (1984), Binggeli et al. (1987), Bothun \& Mould (1988),
Gallagher \& Hunter (1989), Ferguson (1992), Binggeli et al. (1993),
Conselice et al. (2001), Ferrarese et al. (2012), and many other authors. 
In general, early-type gas-deficient galaxies (E, S0, dSp) concentrate
at the cluster centre to a much greater degree than spiral and irregular
gas-rich galaxies. Schindler et al. (1999) analysed the distribution of the
X-ray emission in the Virgo cluster and estimated the mass of the hot
intracluster gas to be three times the integral mass of the galaxies. The
Sunyaev-Zeldovich effect measurements (Planck Collaboration, 2016) confirmed the
presence of the hot intracluster gas with a total mass of about $1.5 \times
10^{14} M_\odot$ being an order of magnitude higher than the integral stellar
mass. Peripheral galaxies lose their gas interacting with the intracluster gas
medium while falling towards the cluster centre. The loss of gas causes the
suppression of the star formation process in galaxies along with their
morphological transformation.

\begin{figure} 
\centering
\includegraphics[width=\hsize]{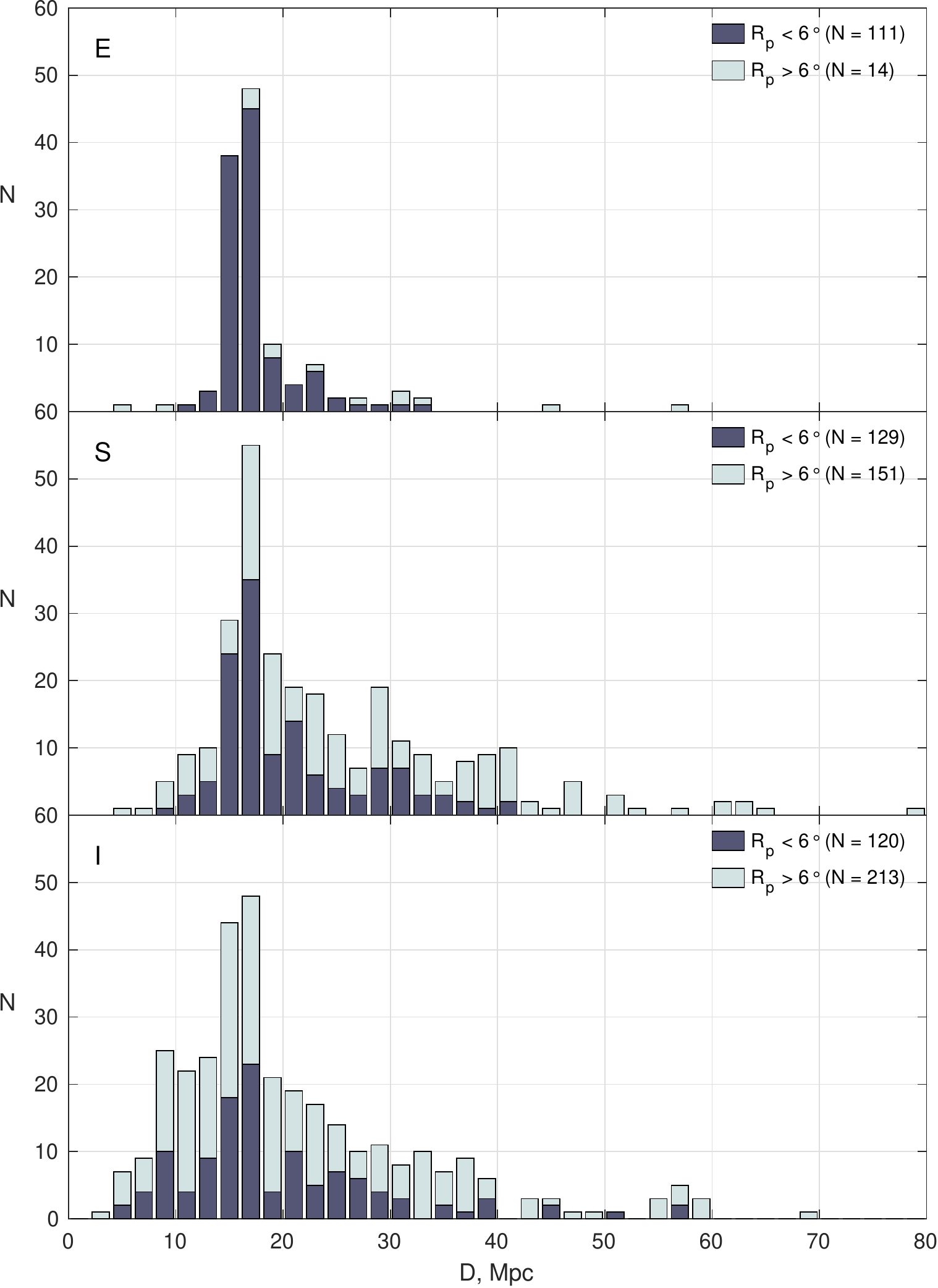}
\caption{Distribution of galaxies of different morphological types 
 in the Virgo cluster area by their
distance $D$. The galaxies within the $6^\circ$ core are marked by a darker colour.}
\label{kkk19:fig07}
\end{figure}

The four panels of Fig.~\ref{kkk19:fig05} reproduce the distribution of all the
galaxies with $V_{LG} < 2600$ km/s, as well as early-type galaxies 
(E, S0, dSph), spiral galaxies, and irregular dwarfs (I, Im, BCD) among them.
Their distribution is smoothed with a window of $0.75^\circ$. As is clear from
these data, the major part of early type galaxies reside within the virial
radius $R_p = 6.0^\circ$. Half of the spiral galaxies lie outside the virial
zone, and many of them are associated with background $W$ cloud. The
distribution of S galaxies is clumpy inside the virial radius, pointing out that
this sub-system has not yet been dynamically relaxed, as noticed by Conselice
et al. (2001). Many late-type dwarf galaxies are concentrated in the background $W$ and $M$ clouds, while a significant
number of them reside in the Virgo Southern Extension region left of the virial
zone. Tully \& Shaya (1984) considered this region as the main channel providing
the virial core of the Virgo cluster with extrinsic galaxies.

The evidence of this assumption can be seen in Fig.~\ref{kkk19:fig06},
illustrating the radial velocities of galaxies in the VirgoSE zone and the
behaviour of the running median with a window of $2^\circ$. The characteristic
radial velocity of galaxies grows from 1100~km~s$^{-1}$ at the left (southern)
border of the considered region ($SGL = 117^\circ$) up to 1400~km~s$^{-1}$ at
the virial boundary of the cluster ($SGL = 109^\circ$). The majority of these
galaxies are located between the observer and the Virgo cluster (see the
upper-left panel of Fig.~\ref{kkk19:fig03}). While falling along the VirgoSE
filament towards the cluster centre, the galaxies with Virgocentric
distances within a range of ($2-4$)~Mpc get an additional line-of-sight
velocity component of $\sim300$~km~s$^{-1}$ that agrees with the results of
N-body simulations (Klypin et al. 2003). As far as we know, this is the first
direct observational evidence of the case that baryonic matter moves along a
cosmic filament toward a cluster as a node of the large-scale structure.

The mean radial velocity $\langle{}V_{LG}\rangle$ and the radial velocity
dispersion $\sigma_V$ of E, S, and I galaxies with distances in a ($12-22$)~Mpc
range are given in Table~\ref{kkk19:tab06}.

\begin{table}
\caption{Mean radial velocity $\langle{}V_{LG}\rangle$ and radial velocity
dispersion $\sigma_V$ for the different galaxy populations within a ($12-22$)~Mpc
distance range.}\label{kkk19:tab06}
\centering
\begin{tabular}{lrrr}\hline\hline
Type   &  N   &$\langle{}V_{LG}\rangle$& $\sigma_V$ \\
       &      &      km\,s$^{-1}$      &km\,s$^{-1}$\\
\hline
E      & 103  &         1039           &     516    \\
S      & 137  &          993           &     674    \\
I      & 158  &         1158           &     676    \\
\hline
All    & 398  &         1070           &     638    \\
\hline
\end{tabular}
\end{table} 

\begin{figure} 
\centering
\includegraphics[width=\hsize]{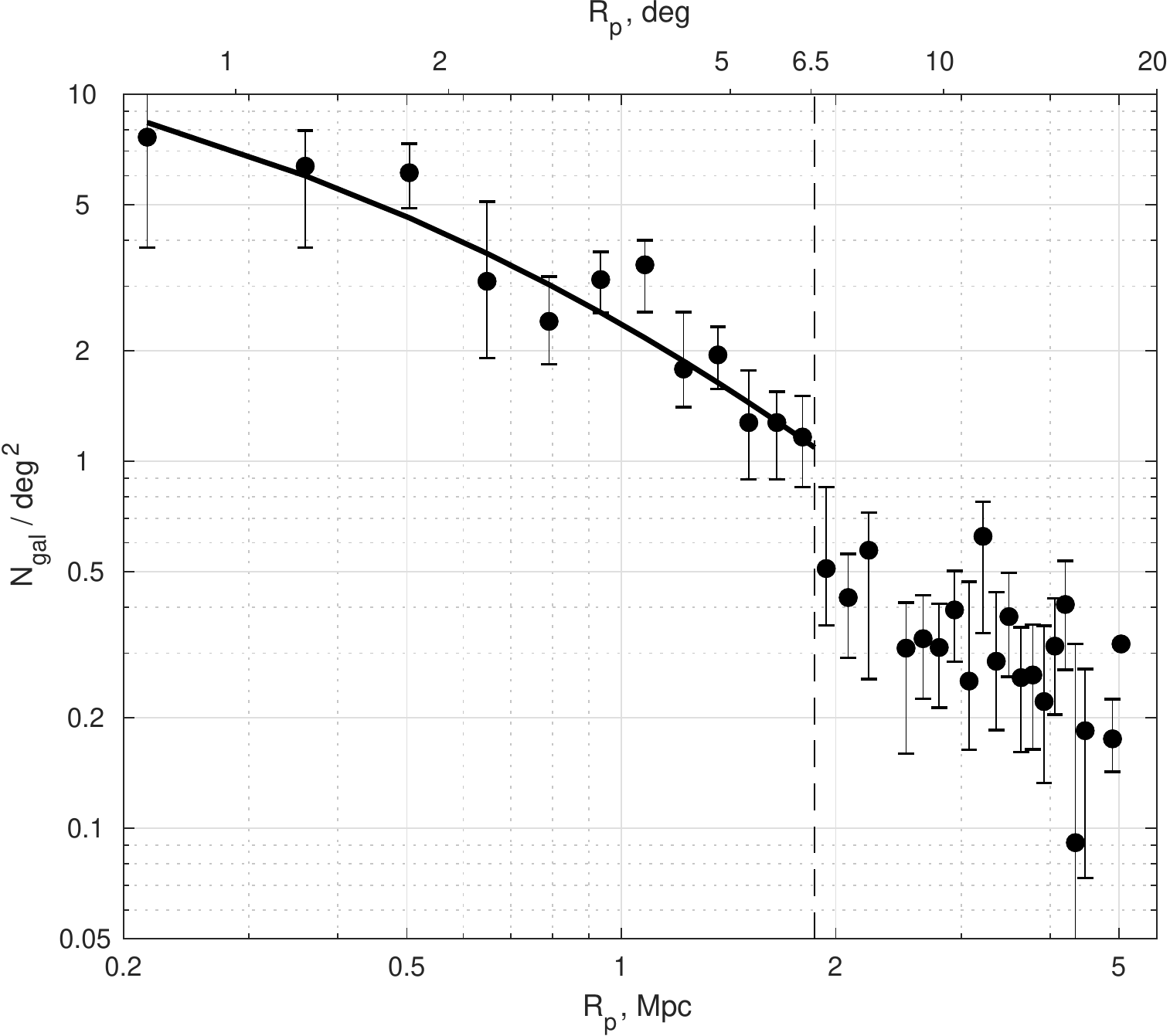}
\caption{Surface density profile of Virgo galaxies as a function of projected
radius $R_p$ given in degrees (upper scale) and Mpc (lower scale). The solid
line corresponds to the standard projected NFW profile of the dark halo. }
\label{kkk19:fig08}
\end{figure}

The distribution of galaxies of different morphological types by their distance
$D$ is shown in three panels of Fig.~\ref{kkk19:fig07}. In these
histograms, galaxies located inside the virial radius of $6^\circ$ are marked
with the darker colour. We can make several conclusions analysing the presented
data: a) Early-type galaxies show a sharp peak within $D = (14-18)$~Mpc distance
range, with a tail at $D = (20-30)$~Mpc region caused by $W$ and $M$ clouds
members.
b) Spiral galaxies inside $R_p = 6.0^\circ$ hit a peak within $(14-18)$~Mpc
too, but the most spirals lie substantially further away than
the Virgo cluster. Among the spirals residing within the $6^\circ$ radius, only half have distances that are typical for Virgo.
c) Of late-type dwarf galaxies, only 1/3 are associated with the
virial core of the cluster; the main population of gas-rich dwarfs lies in
the background or foreground relative to the cluster.

\section{Density profiles and total mass of the cluster}
The distribution of surface density (per square degree) of Virgo galaxies along
the projected radius $R_p$ of the cluster is presented in the logarithmic scale
in Fig.~\ref{kkk19:fig08}. To plot this density profile, we used only the
galaxies with distances in a ($12-22$)~Mpc range. The solid line corresponds to
the standard projected Navarro-Frenk-White (NFW) profile of the dark halo (Navarro et al. 1997,
Bartelmann 1996) with a fitted scale radius $R_s = 4.3^\circ$ or 1.24~Mpc in
projection. One can notice that there is an abrupt edge at $R_{edge} =
6.5^\circ$ that is inferred to correspond to the projected virial radius. Using
the data listed in Table~\ref{kkk19:tab01}, we determined the mean,
$\langle{}R_p\rangle$, and the median for projected distances of galaxies from
M87. The data are presented in Table~\ref{kkk19:tab03} for wide and narrow
ranges of distances: ($12-22$)~Mpc and ($14-20$)~Mpc. We can see that the
$\langle{}R_p\rangle$ uncertainty caused by various choices of cluster members is
not large and does not differ significantly with statistical error ($\sim5$\,\%).

\begin{table}
\caption{Median and average separation of galaxies from M87.}\label{kkk19:tab03}
\centering
\begin{tabular}{cccc}\hline\hline
   D    &$N_D$&    median     &  $\langle R_p\rangle$  \\
 (Mpc)  &     &               &                        \\
\hline
12--22  & 398 & 4.74$^{\circ}$& 5.91$\pm$0.20$^{\circ}$\\
14--20  & 316 & 4.36$^{\circ}$& 5.69$\pm$0.23$^{\circ}$\\
\hline
\end{tabular}
\end{table}

According to Tully (2015), the virial mass of a cluster $M_{VIR}$ is expressed
via the radial velocity dispersion $\sigma_p$ and the projected radius $R_g$ as

\begin{equation}\label{kkk19:eq01} 
M_{VIR} = (\alpha \times \pi/2)G^{-1} \times \sigma_p^2 \times R_g,
\end{equation}

where G is gravitational constant, projected radius $R_g$ is the mean harmonic
distance for all the pairs, and the dimensionless parameter $\alpha\simeq 2.5\pm0.1$
accounts for the conversion of radial velocity dispersion into 3D velocity
dispersion under assumption of weak anisotropy. It should be noted, however, 
that the mean harmonic distance between galaxies is not a robust characteristic 
of a cluster radius due to projection effects. Assuming the mean distance of the 
cluster 16.5~Mpc, radial velocity dispersion $\sigma_p = 638$~km~s$^{-1}$ (see 
below) and $R_g$ in the range between $\langle{}R_p\rangle = 5.7^\circ$ and
$R_{edge} = 6.5^\circ,$ we get the following virial mass estimation for the Virgo cluster:

\begin{equation}\label{kkk19:eq02} 
M_{VIR} = (6.0 - 6.8) \times 10^{14} M_\odot.
\end{equation}


\begin{figure*} 
\centering
\resizebox{0.95\hsize}{!}
{\includegraphics{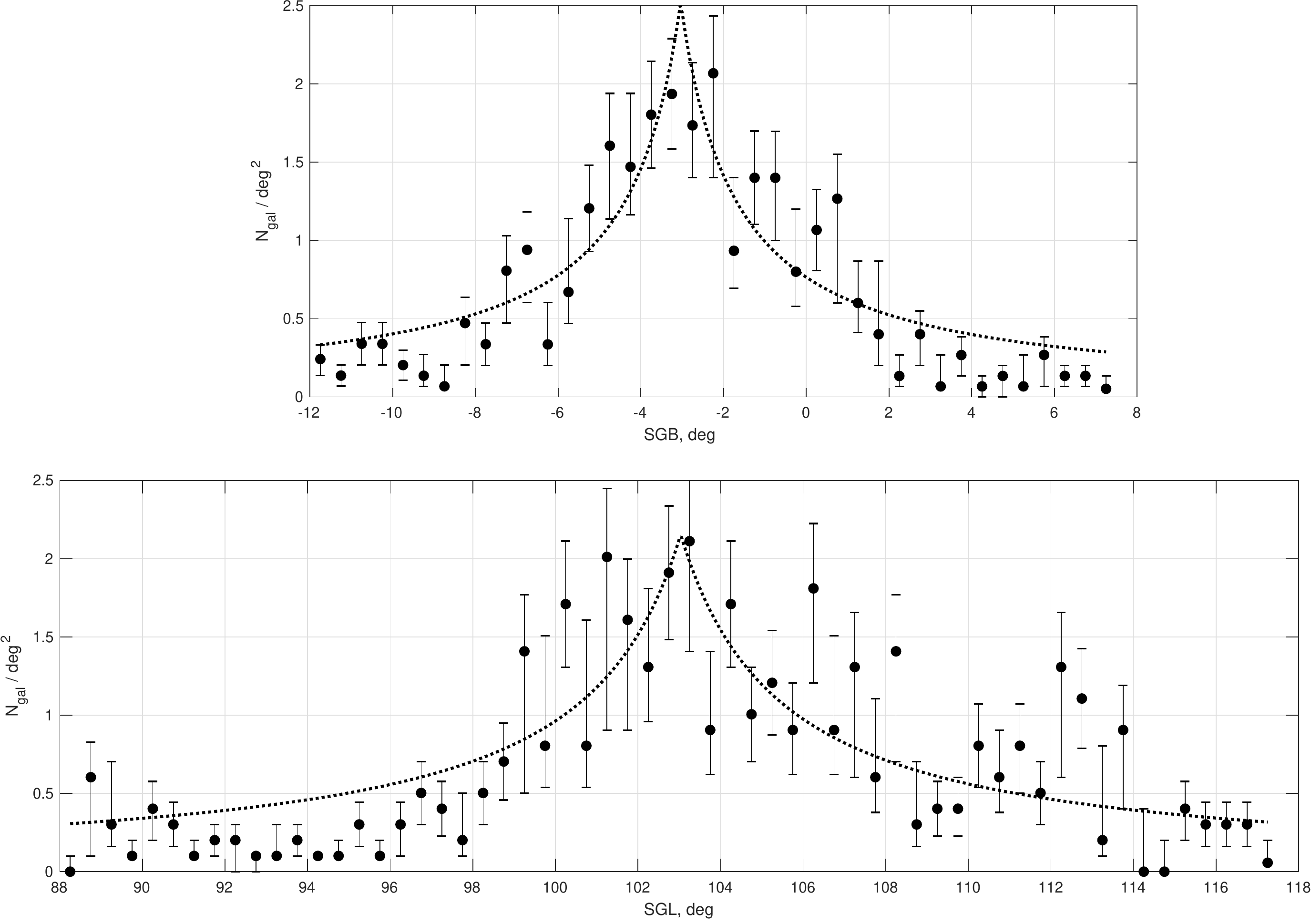}}
\caption{Standard one-dimensional NFW profile for Virgo galaxy counts with
estimated parameters and 95\% confidence intervals.}
\label{kkk19:fig09}
\end{figure*}

An elegant way of determining the mass of a spherically symmetrical cluster was
proposed by Ambartsumian (1939). The potential energy of a system $U$ is
expressed from the density distribution $F(y)$ projected into a line crossing
the cluster centre as

\begin{equation}\label{kkk19:eq03} 
U = G \int^{+\infty}_{-\infty} [F(y)]^2\,dy.
\end{equation}

The problem of finding $U$ can be reduced to galaxy number counts in parallel stripes of
$\Delta{}y$ width. Substituting Eq.~(\ref{kkk19:eq03}) in the virial theorem $2T +
U = 0$ where $T$ is kinetic energy, we obtain the expression for the total mass
of the cluster

\begin{equation}\label{kkk19:eq04} 
M_{VIR} = \nu \alpha \times G^{-1} \times \sigma_p^2 \times \Delta{}y.
\end{equation}

The dimensionless factor

\begin{equation}\label{kkk19:eq05} 
\nu = \frac{(\Sigma{}n_i)^2}{\Sigma{}n_i^2}
\end{equation}

is governed by the galaxy number counts, $n_i$, in parallel stripes of
$\Delta{}y$ width, and another dimensionless factor $\alpha \simeq 2.5$, like in
Eq.~(\ref{kkk19:eq01}), accounts for conversion of radial velocity dispersion into
the full velocity dispersion. 

\begin{table}
\caption{Galaxy counts in strips.}\label{kkk19:tab04}
\centering
\begin{tabular}{cccc}\hline\hline
Strips      &                        &$\nu$&$M$\\
&           &      &($10^{14}\times M_{\odot}$)\\
\hline
Along SGL,  &$\Delta y = 0.5^{\circ}$&19.5 &6.6\\
            &$\Delta y = 1.0^{\circ}$&10.9 &7.4\\
Along SGB,  &$\Delta y = 0.5^{\circ}$&16.1 &5.4\\
            &$\Delta y = 1.0^{\circ}$& 8.4 &5.7\\
\hline
\end{tabular}
\end{table}

The first galaxy counts in stripes, arranged parallel and perpendicular to the
declination, yielded $6.9 \times 10^{14} M_\odot$ and $9.3 \times 10^{14}
M_\odot$, respectively, for the mass of the Virgo cluster (Karachentsev 1965).
We repeated galaxy counts in the Virgo cluster within a $(12-22)$~Mpc
distance range. The counts were performed in stripes of $\Delta{}y = 0.5^\circ$
and $1^\circ$ width, oriented along SGB and SGL. The results are shown in the
two panels of Fig.~\ref{kkk19:fig09}. The obtained values of $\nu$ factor and
the total mass of the cluster are presented in Table~\ref{kkk19:tab04}. The
difference in counts made along SGL and SGB are quite predictable since the Virgo
cluster is non-spherical and elongated by the LSC equator. The mean value of
mass from four estimates is

\begin{equation}\label{kkk19:eq06} 
M = (6.3\pm0.9) \times 10^{14} M_\odot,
\end{equation}

in good agreement with the mass estimation obtained in Eq.~(\ref{kkk19:eq02}) from the
projected radius. The uncertainty of the mass includes the uncertainty of radial
velocity dispersion.

Assuming the standard NFW spatial density profile (Navarro et al. 1997) for the
dark halo of the cluster as

\begin{equation}\label{kkk19:eq07} 
D(x) = D_c/x \times (1+x)^2,
\end{equation}

where $D_c$ is a characteristic density and $x = r/r_s$ is a dimensionless
distance expressed in units of a scale radius $r_s$, we obtain a simple 
equation for density function projected onto the axis:

\begin{equation}\label{kkk19:eq08} 
F(y) = \frac{F_0}{1+y},
\end{equation}

as, by definition, 

\begin{equation}\label{kkk19:eq09} 
F'(x) = -2\pi{}x \times D(x).
\end{equation}

\begin{figure} 
\centering
\includegraphics[width=\hsize]{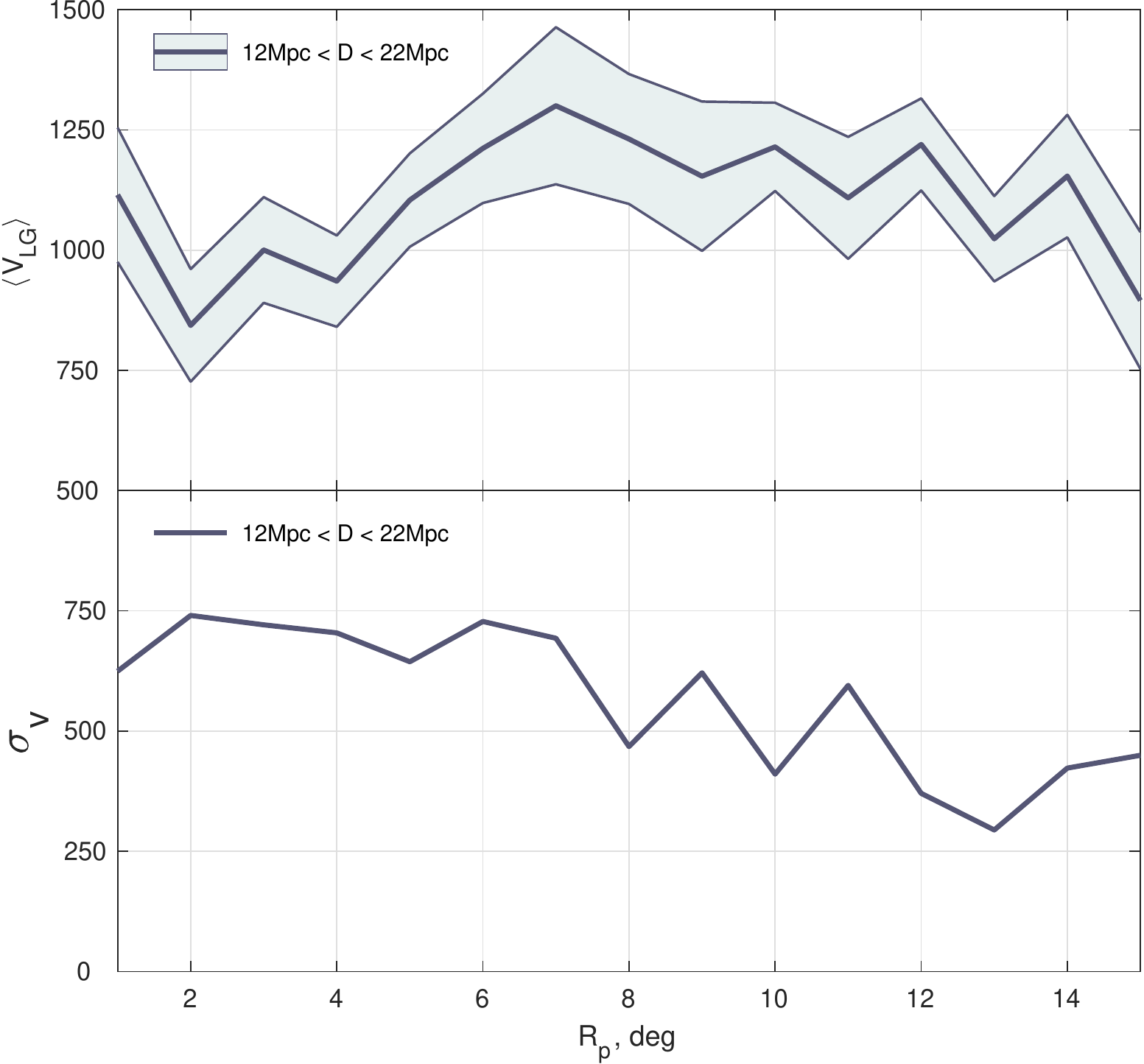}
\caption{Mean radial velocity $\langle{}V_{LG}\rangle$ and radial velocity
dispersion $\sigma_V$ of Virgo galaxies as a function of projected radius.}
\label{kkk19:fig10}
\end{figure}

This standard 1D NFW profile is represented in two panels of
Fig.~\ref{kkk19:fig09} with dotted hyperbola. As we can see, the standard
density profile looks much sharper than the real profile of galaxy counts along
$SGB$ and $SGL$.
The Kolmogorov-Smirnov test for goodness-of-fit gives the evidence of
non-normally distributed residuals with $p$-values of $3.8 \times 10^{-5}$ and
$1.4 \times 10^{-4}$ for galaxy counts along $SGL$ and $SGB$, respectively,
suggesting
a very weak agreement between the observed distribution of galaxies 
and the expected density distribution. In contrast to this, McLaughlin 
(1999) found that the 'universal' NFW model by Navarro et al. (1997)
traces the galaxy distribution in the Virgo cluster very well, yielding
its virial mass of $(4.2\pm0.5) \times 10^{14} M_\odot$. 

We note that Simionescu et al. (2017) investigated X-ray properties of the
Virgo cluster with SUZAKU Key Project and obtained a much lover virial radius,
$r_{200} = 3.36^\circ$, and virial mass estimate, $M_{200} = 1.05 \times
10^{14}M_\odot$, assuming isothermal hot gas to be in hydrostatic equilibrium
within a NFW potential (Navarro et al. 1997).

Finally, it should be stressed that we assume the Virgo cluster is 
virialised. However, as Conselice et al. (2001) demonstrated, only elliptical
galaxies in the Virgo core region form a nearly relaxed subsystem, no other
galaxy populations (spiral or irregular) have characteristics of relaxation.
Also, we assume that the galaxy orbit distribution has a weak anisotropy.
There is, however, ample evidence that many spiral and irregular galaxies move across 
the cluster core in elongated, almost radial orbits. This makes the estimation 
of the total velocity dispersion of galaxies dependent on the morphological
composition of the sample. Obviously, these simplifying assumptions affect
the estimation of the virial mass and its errors. 

If we suppose that only early-type galaxies with their radial velocity
dispersion of 516 km/s are in the relaxed state, the virial mass of the Virgo
cluster turns out to be $4.1 \times 10^{14} M_\odot,$ in close agreement with
the McLaughlin (1999) estimate. Also, replacing the virial theorem $2T + U = 0$ in
the condition of negative total energy of the cluster, $T + U \leqslant 0$,
yields us the estimate of the total cluster mass $M_T \geqslant M_{VIR}/2$.

As mentioned above, the Planck Collaboration (2016) performed a detailed
study of the Virgo cluster through Sunyaev-Zeldovich effect and found the total mass
of warm/hot gas to be $(1.4 - 1.6) \times 10^{14} M_\odot$. Assuming the cosmic
value for the baryon fraction, $f_b = \Omega_b/\Omega_m = 0.1834$, they found that
the total mass of the cluster would be $(7.6 - 8.7) \times 10^{14} M_\odot$ on a
scale up to 2 times larger than the virial radius.

\section{Basic parameters of the Virgo cluster}
The nearest cluster of galaxies in Virgo is qualified as a rather rich one,
though it is not included in the known catalogues by Zwicky et al. (1961) and
Abell (1958) due to its large angular size. The existing substructures around
M87, M49, M86, and other bright galaxies, were extensively distinguished (de
Vaucouleurs 1961, Schindler et al. 1998, Mei et al. 2007, Kim et al. 2014) as being
illustrative of incomplete dynamical relaxation of the Virgo cluster. The
observable infall of late-type galaxies to the virial zone along the VirgoSE
filament also indicates this fact. The Virgo cluster, being the most broadly
studied one, is involved in a comparison of the numerical simulation results with
the observations. This is the reason why it would be instructive to
compile an `identity card' of the cluster containing its principal
observational characteristics.

Two panels of Fig.~\ref{kkk19:fig10} illustrate the dependence of the
cluster mean radial velocity and the radial velocity dispersion on the projected
distance from M87 with a step of $1^\circ$. The analysis introduces galaxies in
a ($12-22$)~Mpc distance range. The
variations of the mean radial velocity seem to be significant and appreciably
exceed the SEM error bars. Besides the virial core, $R_p > 6^\circ$, the
observable variations are caused mainly by the kinetic pattern of the ongoing
infall of galaxies towards the cluster centre. The decreasing trend of radial
velocity dispersion towards the outskirts of the cluster bears evidence of
mainly radial flows of galaxies. The mean value of radial velocity is
$\langle{}V_{LG}\rangle = 1070\pm48$~km~s$^{-1}$ for the virial zone of the
cluster, and the corresponding radial velocity dispersion is
$638\pm35$~km~s$^{-1}$. Assuming $16.5\pm0.2$~Mpc as the cluster distance and
$H_0 = 73$~km~s$^{-1}$~Mpc$^{-1}$ as the local Hubble parameter, we obtain the
value $-135\pm50$~km~s$^{-1}$ as the Virgocentric infall velocity of the Local
Group.

The principal integral characteristics of the Virgo cluster are listed in Table~\ref{kkk19:tab05}. In most cases, the cited errors of mean values are
formal, since there are different underlying factors that affect the estimate of
the mean but are difficult to quantify.\begin{table}
\caption{Average parameters of the Virgo cluster.}\label{kkk19:tab05}
\centering
\begin{tabular}{ccl}\hline\hline
Parameter              &   Derived   & Unit               \\
                       &    value    &                    \\
\hline
$\langle V_{LG}\rangle$&$1070\pm48 $ & km/s               \\
$\sigma_p$             &$ 638\pm35 $ & km/s               \\
$\langle D\rangle$     &$16.5\pm0.2$ & Mpc                \\
$R_g$                  &$ 1.7\pm0.2$ & Mpc                \\
$L_B$                  &$ 1.8\pm0.2$ & $10^{12} L_{\odot}$\\
$L_K$                  &$ 8.6\pm1.1$ & $10^{12} L_{\odot}$\\
$M_{vir}$              &$ 6.3\pm0.9$ & $10^{14} M_{\odot}$\\
\hline
\end{tabular}
\end{table}

\section{Concluding remarks}
The analysis of galaxy motions in the outskirts of the Virgo cluster makes it possible to
measure the radius of the zero-velocity surface, $R_0 = 7.0-7.3$~Mpc
(Karachentsev et al. 2014, Shaya et al. 2017, Kashibadze et al. 2018),
corresponding to the total mass of the Virgo cluster $M_T = (7.4\pm0.9) \times
10^{14}M_\odot$ inside the $R_0$. The numerical simulated trajectories of nearby
galaxies with accurate distance estimates performed by Shaya et al. (2017)
confirmed the obtained estimate of the total mass of the cluster.
The virial mass of the cluster, being determined independently at the scale of
$R_g = 1.7$~Mpc from the internal motions, is nearly the same - $M_{VIR} =
(6.3\pm0.9) \times 10^{14}M_\odot$.
The agreement of internal and external mass estimates within their nominal error
of $\sim15$\,\% attests to the fact that the wide outskirts of the Virgo cluster
between $R_g$ and $R_0$ radii do not contain significant amounts of dark matter.
This observational result seems to be a challenge for certain cosmological
models within the standard $\Lambda$CDM paradigm, affirming that the peaks of dark
matter large-scale distribution are less sharp than the stellar matter peaks
(Fukugita \& Boehringer 2019). The explanation of this controversy requires new
observational and theoretical efforts.
 
\begin{acknowledgements}
 We thank the anonymous referee whose suggestions and comments helped to improve
significantly the presentation of the paper results. The work is supported by
the RFBR grant No. 18--02--00005. We acknowledge the usage of SDSS, PanSTARRS
and ALFALFA surveys, as well as the HyperLeda\footnote{http://leda.univ-lyon1.fr} and
NED\footnote{http://ned.ipac.caltech.edu} databases.
\end{acknowledgements}

\end{document}